\documentclass[aps,prb,preprintnumbers,nofootinbib,floatfix,11pt]{revtex4}

\usepackage{amsfonts,amsmath,amssymb,dsfont,yfonts,graphicx,epsfig,color,yhmath,natbib,mathtools,appendix}
\setcitestyle{super, sort&compress}

\newcommand{\be}{\begin{equation}}
\newcommand{\ee}{\end{equation}}
\newcommand{\ba}{\begin{eqnarray}}
\newcommand{\ea}{\end{eqnarray}}

\DeclareMathOperator{\csch}{csch}

\newcommand{\veryshortarrow}[1][3pt]{\mathrel{%
   \hbox{\rule[\dimexpr\fontdimen22\textfont2-.2pt\relax]{#1}{.4pt}}%
   \mkern-4mu\hbox{\usefont{U}{lasy}{m}{n}\symbol{41}}}}

\makeatletter

\setbox0\hbox{$\xdef\scriptratio{\strip@pt\dimexpr
    \numexpr(\sf@size*65536)/\f@size sp}$}

\newcommand{\scriptveryshortarrow}[1][3pt]{{%
    \hbox{\rule[\scriptratio\dimexpr\fontdimen22\textfont2-.2pt\relax]
               {\scriptratio\dimexpr#1\relax}{\scriptratio\dimexpr.4pt\relax}}%
   \mkern-4mu\hbox{\let\f@size\sf@size\usefont{U}{lasy}{m}{n}\symbol{41}}}}

\makeatother

\begin{document}

\title{The Harmonic Quantum Szil\'ard Engine}
\date{\today}
\author{Paul Davies, Logan Thomas, George Zahariade}

\affiliation{Department of Physics and Beyond: Center for Fundamental Concepts in Science\\
Arizona State University, Tempe, Arizona 85287, USA} 

\begin{abstract}
The Szil\'ard engine is a mechanism (akin to Maxwell's demon) for converting information into energy, which seemingly violates the second law of thermodynamics. Originally a classical thought experiment, it was extended to a quantized treatment by Zurek. Here, we examine a new, elegant model of a quantum Szil\'ard engine by replacing the traditional rigid box with a harmonic potential, extending the scope of the model. Remarkably, almost all calculations are exact. This article is suitable to students, researchers and educators interested in the conceptual links between information, entropy, and quantum measurement.
\end{abstract}

\maketitle

\section{Introduction}

The concept of Maxwell's demon has occupied a key place in the history of physics, sitting at the intersection of thermodynamics, information theory and computing. Originally merely a thought experiment, advances in nanotechnology have recently brought the demon into the experimental domain. In 1929, Leo Szil\'ard sought to simplify Maxwell's original conception with an imaginary device that became known as Szil\'ard's engine.\cite{1929ZPhy...53..840S} It consisted of a single classical particle confined to a finite region of space (let us assume it to be a rectangular box) and coupled to a thermal bath at temperature $T$.  A demon determines the location of the particle, and inserts a moveable barrier in the middle of the box that can slide without friction like a piston, in such a way as to extract useful work from the isothermal expansion of the single-particle ideal gas, without itself expending any energy. In the ideal limit, the amount of work extracted turns out to be $W = k_BT \ln 2$, where $k_B$ is Boltzmann’s constant, this being comparable to the average kinetic energy of a single molecule of gas. As a result of the manipulation, the entropy of the gas decreases by an amount $\Delta S = k_B \ln 2$, corresponding to a Shannon entropy of one bit,\cite{ShannonEntropy} defined as the reduction in uncertainty on inspecting the outcome of a process with probability 1/2, e.g. flipping an unbiased coin. The Szil\'ard engine has thus converted information about the particle's location into useful energy, showing that information can serve as a type of fuel. Several recent experiments have confirmed this prediction by demonstrating the conversion of information into work, both in the classical\cite{KoskiEtAl14,KoskiEtAl15,KoskiReview,SerreliEtAl07,ToyabaEtAl10} and quantum regimes.\cite{CottetEtAl,NaghilooEtAl,CamatiEtAl} In addition, biophysicists have identified many molecular machines that play the role of Maxwell demons in living organisms.\cite{BoelEtAl19}

The apparent violation of the second law of thermodynamics implied by Szil\'ard's engine has been the subject of thorough investigation (see, e.g. Refs.\citenum{ParrondoEtAl} and \citenum{LeffRex90} and references therein for a nice summary). Szil\'ard himself understood that, to operate in a cyclic manner, there would be an entropic cost greater than the thermodynamic profit. That cost may arise from garnering infomation about the atom's position or resetting the engine to its initial state. When the measurement device does not store information, such as in demonless Szil\`ard engines and automated Maxwell's demons,\cite{LeffRex94,DemonlessSzilard,SkordosZurek92} the cost arises from the information acquisition. However, Landauer\cite{Landauer61} and Bennett\cite{Bennett03} later argued that it is possible to make that step reversible (i.e. with zero entropic cost). In that case, in order for the engine to operate in a cyclic manner, the information acquired and stored in a register (the demon's ``brain'') would have to be erased after each cycle to reset the register to a {\it tabula rasa} state. The entropy of erasure is at least equal to $k_B \ln 2$, according to Landauer's principle.\cite{Landauer61} Boyd \& Crutchfield then showed that the compensating entropic burden to pay for the demon's work could be distributed between information acquisition and erasure.\cite{BoydCrutchfield16} The key to balancing the entropic books lies with incorporating the demon into the total physical system. 

Although the foregoing analyses resolve any thermodynamic paradoxes, they suffer from the shortcoming of being entirely classical, which is clearly unrealistic when referring to the position measurement of a single atom. A quantum version of Szil\'ard's original atom-in-a-box model can be realized with a one-dimensional non-relativistic quantum particle of mass $m$ confined by a symmetric potential $V(q)$, $q$ being the particle's position, to a finite region of space. The quantum dynamics are described by the Hamiltonian
\be
\hat{H}=\frac{\hat{p}^2}{2m}+V(\hat{q})\,,
\ee 
where $V(\pm\infty)=\infty$, $p$ is the particle's momentum, and the particle is taken to be in a thermal state. The demon then quasi-statically inserts a thin barrier (modeled by a time-dependent potential) at the center of the well, and performs a strong projective measurement\cite{basdevant2002quantum} on the particle's position to determine whether it is located to the left or right of the partition. With this information, the demon can extract work from the setup. The demon can be considered either as an external ``observer,'' or itself a quantum system with internal states that couples to the atom. In either case, the act of measurement involves an irreversible collapse of the wave function. Transitioning to a quantum treatment opens up some additional foundational issues that have been addressed both theoretically\cite{ElouardEtAl17,AshrafiEtAl20} and experimentally\cite{Pekola15}. Here, however, we sidestep these subtleties and present a basic pedagogical treatment of a quantum Szil\'ard engine that more closely parallels Szilard’s original concept.

The case of the infinite square well potential was studied in detail by Zurek.\cite{2018PhR...755....1Z} 
An elegant model that captures the same essential physical processes is obtained by replacing the box (i.e. the square well potential) with a harmonic oscillator potential, of angular frequency $\omega$, $V(q)=~\frac{1}{2}m\omega^2q^2$, for which the partition function of the particle can be obtained exactly, and all the relevant thermodynamic quantities immediately extracted. The aim of this work is to describe this improved version in detail. Similar work involving Landauer's principle for a quantum particle in a harmonic well can be found in Ref.~\citenum{Klaers19}, though in a context divorced from the Szil\'ard engine. The quantum dynamics involved in the insertion of a barrier was studied in Ref.~\citenum{BaekEtAl16}. Further work on Landauer's principle in the classical, semi-classical, and quantum regimes of the Szil\'ard engine can be found in Ref.~\citenum{AshrafiEtAl20}.
 
In Sec.~\ref{cycle}, we analyze the evolution of the quantum system during one Szil\'ard engine cycle. The quantum particle is prepared in an initial thermal state of the harmonic well (Sec.~\ref{initial}).  Next, a Dirac delta potential barrier is quasi-statically inserted in the middle of the well, thus modifying the quantum state of the particle (Sec.~\ref{barrier}). Then, a quantum position measurement allows the demon to localize the particle on the left or the right side of the barrier (Sec.~\ref{measurement}). The one-particle quantum gas finally expands isothermally, which allows the demon to extract useful work (Sec.~\ref{joule}) and seems to violate the second law of thermodynamics. In Sec.~\ref{secondlaw}, we explain why this is, in fact, not the case when treating the demon itself as part of the system. We end in Sec.~\ref{discussion} with a short discussion. Our treatment should be suitable for advanced undergraduates and researchers interested in the conceptual links between information, entropy, and quantum measurement, or for use in a graduate course on quantum mechanics, statistical mechanics or (quantum) information theory.

\section{Description of the thermodynamic cycle}
\label{cycle}

\subsection{Initial state}
\label{initial}

We begin by characterizing the initial state of the Szil\'ard engine cycle. As mentioned in the introduction, the initial Hamiltonian is
\be
\hat{H}_{\rm in}=\frac{\hat{p}^2}{2m}+\frac{1}{2}m\omega^2 \hat{q}^2\,.
\ee
Consider a quantum particle  in a (mixed) thermal state at temperature $T$, described by the  density matrix 
\be
\hat{\rho}_{\rm in}=\frac{1}{Z_{\rm in}}\sum_{n=0}^\infty e^{-\beta (n+1/2)\hbar\omega}|\psi_n\rangle\langle\psi_n|\,.
\label{densityinit}
\ee
Here the $|\psi_n\rangle$ are the usual eigenstates of the simple harmonic oscillator corresponding to energies $E_n=(n+1/2)\hbar\omega$, and $\beta=\frac{1}{k_BT}$. The partition function $Z_{\rm in}$ is given by
\be
Z_{\rm in}={\rm Tr}\left(e^{-\beta\hat{H}_{\rm in}}\right)=\frac{e^{-\frac{1}{2}\beta\hbar\omega}}{1-e^{-\beta\hbar\omega}}=\frac{1}{2}\csch\left(\frac{\beta\hbar\omega}{2}\right)\,.
\ee
This quantity is extremely useful since it permits the computation of all relevant thermodynamic quantities.\cite{ThermoBook} The Helmholtz free energy, for instance, is simply
\be
A_{\rm in}= -\frac{1}{\beta}\ln Z_{\rm in}= \frac{1}{\beta}\ln\left[2\sinh\left(\frac{\beta\hbar\omega}{2}\right)\right]\,,
\ee
while the average energy is seen to be
\be
E_{\rm in}=-\frac{1}{Z_{\rm in}}\frac{dZ_{\rm in}}{d\beta}=\frac{1}{2}\hbar\omega\coth\left(\frac{\beta\hbar\omega}{2}\right)\,.
\ee
In the low temperature limit ($\beta\hbar\omega\gg 1$), this formula yields an average energy of $\frac{1}{2}\hbar\omega$, as expected since the particle settles in its ground state at zero temperature. However, in the high temperature limit ($\beta\hbar\omega\ll 1$), the average energy is seen to be $k_B T$. This is consistent with the equipartition theorem which ascribes an energy of $\frac{1}{2}k_BT$ to every quadratic Hamiltonian degree of freedom in the high-temperature, classical, limit; here, there are two such degrees of freedom, $p$ and $q$, since we are doing thermodynamics in a harmonic well, as opposed to a rectangular box. We can also calculate the entropy of the system
\be
S_{\rm in}=-k_B {\rm Tr}\left(\hat{\rho}_{\rm in}\ln \hat{\rho}_{\rm in}\right)= -\frac{dA_{\rm in}}{dT}= k_B\left\{\frac{\beta\hbar\omega}{2}\coth\left(\frac{\beta\hbar\omega}{2}\right)-\ln\left[2\sinh\left(\frac{\beta\hbar\omega}{2}\right)\right]\right\}\,.
\ee
As expected from the third law of thermodynamics, in the low temperature limit, the entropy is seen to vanish. In the high temperature limit, the entropy is $S_{\rm in}=k_B \ln\left(\frac{k_BT}{\hbar\omega}\right)$.

\subsection{Barrier insertion}
\label{barrier}

 In order to localize the particle to the left or to the right of the potential well, an infinitely thin potential barrier is quasi-statically inserted at $q=0$ so that the Hamiltonian becomes $\hat{H}_{\rm bar}(t)=\frac{\hat{p}^2}{2m}+\frac{1}{2}m\omega^2 \hat{q}^2+\alpha(t)\delta(\hat{q})$. The strength of this delta function barrier is given by the time-dependent function $\alpha(t)$ which satisfies $\alpha(-\infty)=0$ and $\alpha(+\infty)=\infty$,  the latter condition ensuring that the particle is unable to tunnel through the barrier at late times. We also impose the slowness condition $|\dot{\alpha}/\alpha|\ll \omega$, which allows us to treat the evolution of the wave function adiabatically.\cite{BaekEtAl16} Notice that this hypothesis ensures that the system is in thermal equilibrium with the bath at temperature $T$ at every instant. 
During this quasi-static evolution, the eigenstates $|\psi_n\rangle$ of $\hat{H}_{\rm in}$ slowly vary but are instantaneous eigenstates $|\psi_n^\alpha\rangle$ of $\hat{H}_{\rm bar}$. Therefore the density matrix~\eqref{densityinit} also changes. We are interested in its final expression, after the barrier has been fully inserted, i.e. at $t=+\infty$.

\begin{figure}
      \includegraphics[width=\textwidth]{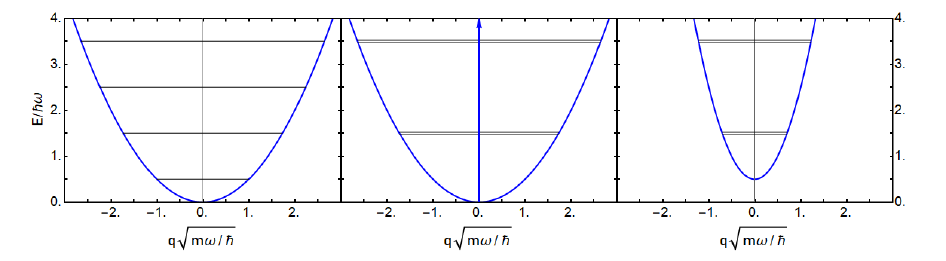}
 \caption{Left: The simple harmonic oscillator potential (blue curve), with energy levels represented by the horizontal lines. Center: The simple harmonic oscillator potential plus the delta function partition in the limit $\alpha\rightarrow\infty$. Energy levels are represented by horizontal lines, with double lines representing the two-fold degeneracy. Right: The energy levels of the harmonic oscillator with a delta function are the same as those of a harmonic oscillator with frequency $2\omega$, shifted up by an energy $\hbar\omega/2$, with two-fold degeneracy at each energy level.}
\label{potential}
\end{figure}

We first need to compute the instantaneous eigenstates $|\psi_n^\alpha\rangle$ and eigenvalues $E^{\alpha}_n$. Their wave functions $\Psi_n^\alpha$ are solutions of the Schr\"odinger equation
\begin{figure}[t]
      \includegraphics[width=0.4\textwidth]{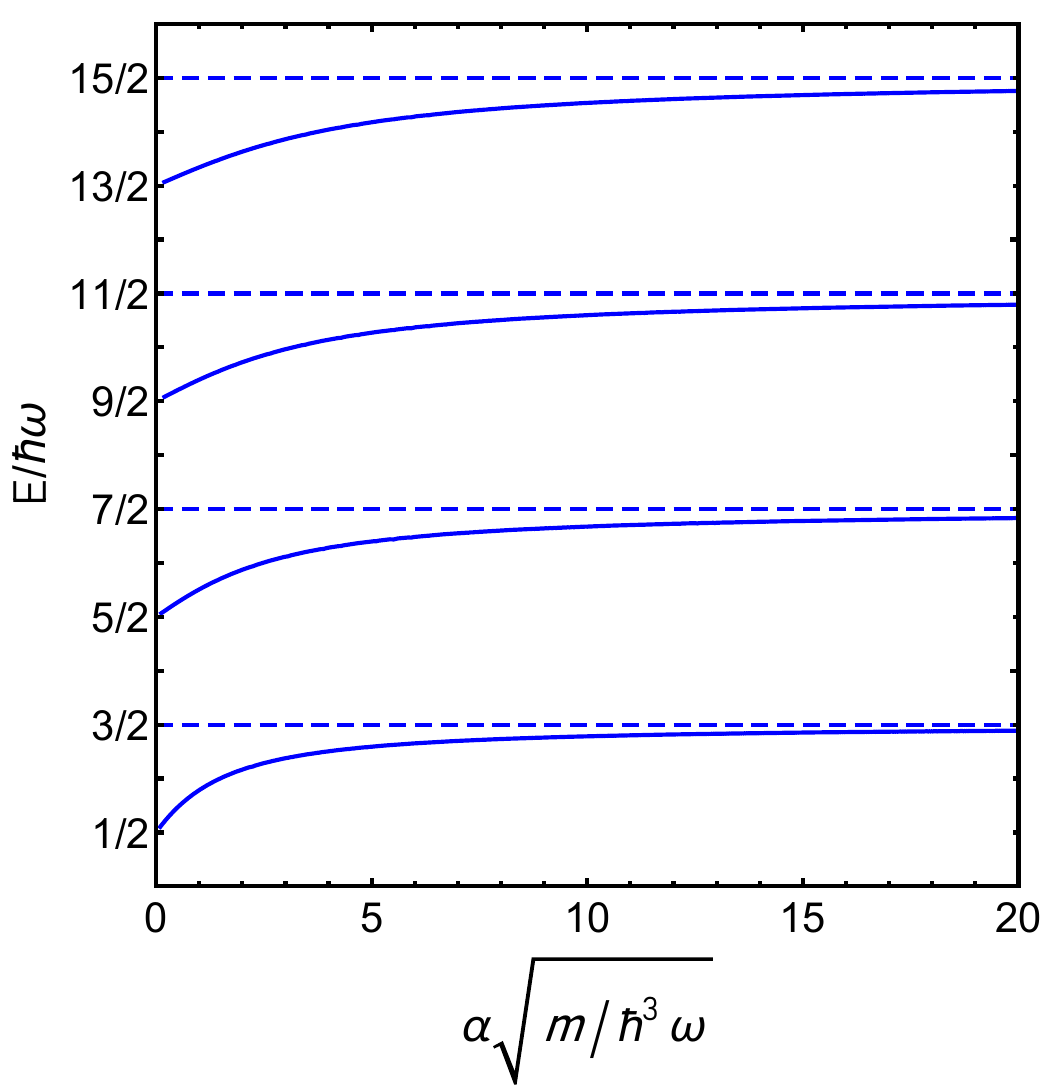}
 \caption{Graphical solution to the implicit equation \eqref{quantization} relating $E$ and $\alpha$; the solid blue curves represent the energy levels of the even modes for a given value of $\alpha$. The dashed blue lines show the energy levels of the odd modes.}
\label{graphsol}
\end{figure}
\be
-\frac{\hbar^2}{2m}\frac{d^2}{dq^2}\Psi^\alpha(q)+\frac{1}{2}m\omega^2q^2\Psi^\alpha(q)+\alpha\delta(q)\Psi^\alpha(0)=E^\alpha\Psi^\alpha(q)\,.
\label{instchrodinger}
\ee
Notice that for odd $n$, the initial eigenstates $|\psi_n\rangle$ (with wave functions $\Psi_n$) and their energy eigenvalues $E_n=(n+1/2)\hbar\omega$ will not be affected by the presence of the barrier since the corresponding wave functions are odd and vanish at $q=0$. (This can be seen by integrating~\eqref{instchrodinger} from $-\epsilon$ to $+\epsilon$ and taking the limit of vanishing $\epsilon$.) In other words, $|\psi_{2k+1}^\alpha\rangle = |\psi_{2k+1}\rangle$ and $E^\alpha_{2k+1}=E_{2k+1}$ for all $\alpha$ (i.e. at all times), which gives the odd solutions, $\Psi_{2k+1}^\alpha=\Psi_{2k+1}$, to Eq.~\eqref{instchrodinger}.
The even solutions, however, require a little work (see Appendix A). We find that, for a particular value of $\alpha$, the energy levels of even modes are given implicitly by the equation
\be
\alpha\sqrt{\frac{m}{\hbar^3\omega}} = -2\frac{\Gamma\left(\frac{3}{4}-\frac{E}{2\hbar\omega}\right)}{\Gamma\left(\frac{1}{4}-\frac{E}{2\hbar\omega}\right)}.
\label{quantization}
\ee
It is easy to solve this equation graphically (see Fig.~\eqref{graphsol}), noticing that as $\alpha$ increases from 0 to $\infty$, the quantized energy levels gradually increase from values $(2k+1/2)\hbar\omega$ to $(2k+3/2)\hbar\omega$. Thus, for even $n$, the initial eigenstates $|\psi_{n}\rangle$ and their energy eigenvalues $E_{n}$ will be affected by the presence of the barrier. More precisely, as $\alpha\gg\sqrt{\hbar^3\omega/m}$, $E^\alpha_{2k}= E_{2k+1}$ and the (even) wave functions $\Psi^\alpha_{2k}$ will satisfy, 
\be
\Psi^\alpha_{2k}(q)=
\begin{cases}
\phantom{-}\Psi_{2k+1}(q),\quad\text{for $q>0$}\,,\\
-\Psi_{2k+1}(q), \quad\text{for $q<0$}\,,
\end{cases}
\ee
This provides the even solutions to Eq.~\eqref{instchrodinger} in the limit $\alpha\rightarrow\infty$ and completes the analysis of the eigenstates $|\psi_n^\infty\rangle$ of the Hamiltonian after the barrier has been inserted. Notice that each energy eigenstate is now degenerate. This effectively produces the spectrum of a simple harmonic oscillator of frequency $2\omega$, with the bottom of the well shifted up by $\hbar\omega/2$, and with each energy level having degeneracy 2 (see Fig.~\ref{potential}).

We are now in a position to write the final state of the thermal density matrix~\eqref{densityinit} (after the barrier has been fully inserted) as
\be
\hat{\rho}_\perp =
\frac{1}{Z_\perp}\sum_{n=0}^\infty e^{-\beta\hbar\omega(2n+3/2)}\left(|\psi^\infty_{2n}\rangle\langle\psi^\infty_{2n}|+|\psi^\infty_{2n+1}\rangle\langle\psi^\infty_{2n+1}|\right)\,.
\ee
Here, the partition function after barrier insertion is given by
\be
Z_\perp={\rm Tr}\left(e^{-\beta \hat{H}_{\rm bar}(\infty)}\right)=\frac{2e^{-\frac{3}{2}\beta\hbar\omega}}{1-e^{-2\beta\hbar\omega}}=e^{-\frac{\beta\hbar\omega}{2}}\csch\left(\beta\hbar\omega\right)\,.
\ee

As in the previous section, we can use the partition function to compute all the relevant thermodynamic quantities. The Helmholtz free energy and the average energy are respectively seen to be
\be
A_{\perp}= -\frac{1}{\beta}\ln Z_{\perp}= \frac{1}{\beta}\ln\left[\sinh\left(\beta\hbar\omega\right)\right]+\frac{\hbar\omega}{2}\,,
\ee
and
\be
E_{\perp}=-\frac{1}{Z_{\perp}}\frac{dZ_{\perp}}{d\beta}=\hbar\omega\coth\left(\beta\hbar\omega\right)+\frac{1}{2}\hbar\omega\,.
\ee
Notice that the insertion of the barrier cannot be done at zero energy cost, since the free energy of the system changes by an amount $A_{\perp}-A_{\rm in}=\frac{1}{2}\hbar\omega+\frac{1}{\beta}\ln\left[\cosh\left(\frac{\beta\hbar\omega}{2}\right)\right]\sim \hbar\omega$; the demon has to provide this energy to the system in the form of work. The entropy of the system is now
\be
S_{\perp}=-k_B {\rm Tr}\left(\hat{\rho}_{\perp}\ln \hat{\rho}_{\perp}\right)= -\frac{dA_{\perp}}{dT}= k_B\left\{\beta\hbar\omega\coth\left(\beta\hbar\omega\right)-\ln\left[\sinh\left(\beta\hbar\omega\right)\right]\right\}\,.
\ee
This is readily understood because the adiabatic insertion of the barrier changes the spectrum to that of a simple harmonic oscillator with double the frequency (hence $\omega \rightarrow 2 \omega$), with each energy level having degeneracy 2. Notice that, in the high temperature limit, $S_{\rm in}=S_\perp$, and the effect of the insertion of the barrier is therefore negligible.

\subsection{Quantum measurement}
\label{measurement}

\begin{figure}[t]
\includegraphics[width=\textwidth]{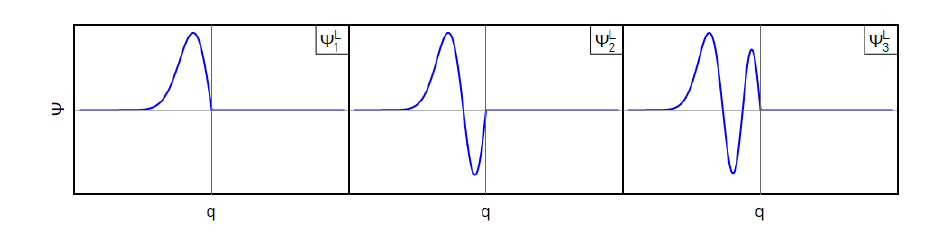}
\caption{The wave functions for the left ground state, $\Psi^L_1$, and first two left excited states, $\Psi^L_2$ and $\Psi^L_3$. The companion right state wave functions can be obtained by flipping the wave functions across the $q=0$ axis.}
\label{elln}
\end{figure}

Before discussing the effect of a strong (projective) quantum measurement on the system,~\cite{endnote0} it is useful to introduce the ``left'' and ``right'' eigenstates
\ba
|L_n\rangle&=& \frac{1}{\sqrt{2}}\left(|\psi_{2n}^\infty\rangle-|\psi_{2n+1}^\infty\rangle\right)\,,\\
|R_n\rangle&=& \frac{1}{\sqrt{2}}\left(|\psi_{2n}^\infty\rangle+|\psi_{2n+1}^\infty\rangle\right)\,,
\ea
and rewrite the density matrix after barrier insertion as
\be
\hat{\rho}_\perp =
\frac{1}{Z_\perp}\sum_{n=0}^\infty e^{-\beta\hbar\omega(2n+3/2)}\left(|L_n\rangle\langle L_n|+|R_n\rangle\langle R_n|\right)\,.
\label{densityfin}
\ee
The wave functions of the states $|L_n\rangle$ and $|R_n\rangle$ are non-zero only over the negative and positive real axis respectively, and they can thus describe a quantum particle located on the left or right side of the potential well (see Fig.~\ref{elln}).

We are now interested in how the demon can determine whether the particle is located to the left of the barrier or to the right. This can be decided by projecting the state of the particle onto one of the two eigenspaces of the observable~\cite{2018PhR...755....1Z}
\be
\hat{\Pi}=\sum_{n=0}^\infty \left(|L_n\rangle\langle L_n|-|R_n\rangle\langle R_n|\right)\,.
\label{observable}
\ee
The associated projectors are
\ba
\hat{P}_L=\sum_{n=0}^\infty |L_n\rangle\langle L_n|\,,\\
\hat{P}_R=\sum_{n=0}^\infty |R_n\rangle\langle R_n|\,,
\ea
and the effect of a projective measurement of $\hat{\Pi}$ on the state of the particle reduces the density matrix $\hat{\rho}_\perp$ to
\ba
\hat{\rho}_{L,R}=\frac{\hat{P}_{L,R}\,\hat{\rho}_\perp\,\hat{P}_{L,R}}{\text{Tr}(\hat{P}_{L,R}\,\hat{\rho}_\perp)}\,.
\ea
More precisely we get that
\ba
\hat{\rho}_L&=&\frac{1}{Z_L}\sum_{n=0}^\infty e^{-\beta\hbar\omega(2n+3/2)}|L_n\rangle\langle L_n|\,,\\
\hat{\rho}_R&=&\frac{1}{Z_R}\sum_{n=0}^\infty e^{-\beta\hbar\omega(2n+3/2)}|R_n\rangle\langle R_n|\,,
\ea
where the partition functions $Z_L$ and $Z_R$ are defined as partial traces over the Hilbert spaces spanned by the left and right eigenstates respectively:
\ba
Z_{L,R}=\frac{1}{2}Z_\perp=\frac{e^{-\frac{3}{2}\beta\hbar\omega}}{1-e^{-2\beta\hbar\omega}}=\frac{1}{2}e^{-\frac{\beta\hbar\omega}{2}}\csch\left(\beta\hbar\omega\right)\,.
\label{ZLR}
\ea

Again, we can use this partition function to compute all the relevant thermodynamic quantities after the particle is measured to be on the left or on the right side of the harmonic well. The Helmholtz free energy and the average energy are respectively given by
\be
A_{L,R}= -\frac{1}{\beta}\ln Z_{L,R}= \frac{1}{\beta}\ln\left[\sinh\left(\beta\hbar\omega\right)\right]+\frac{\hbar\omega}{2}+\frac{1}{\beta}\ln 2\,,
\ee
and
\be
E_{L,R}=-\frac{1}{Z_{L,R}}\frac{dZ_{L,R}}{d\beta}=\hbar\omega\coth\left(\beta\hbar\omega\right)+\frac{1}{2}\hbar\omega\,.
\ee
Notice that the measurement process leaves the average energy unchanged. It does however change the entropy of the system in a crucial way. Indeed, the entropy after ``collapse of the wave function'' to the left or to the right of the harmonic potential well is seen to be
\be
S_{L,R}=-k_B {\rm Tr}\left( \hat{\rho}_{L,R} \ln \hat{\rho}_{L,R}\right) =-\frac{dA_{L,R}}{dT}= k_B\left\{\beta\hbar\omega\coth\left(\beta\hbar\omega\right)-\ln\left[\sinh\left(\beta\hbar\omega\right)\right]\right\}-k_B\ln 2\,.
\ee 
Thus the entropy of the system decreases by exactly 1 bit ($\Delta S=-k_B\ln 2$) as a consequence of the demon acquiring information on the localization of the particle; the irreversible quantum measurement process increases the free energy of the system by a quantity $A_{L,R}-A_\perp=k_B T\ln 2$. As will be shown in the next subsection, this surplus can then be converted into useful work by the demon.

\subsection{Quantum isothermal expansion}
\label{joule}

Once the quantum system has been projected onto one side of the partition, say the right side, the density matrix is $\hat{\rho}_R$ and the barrier is subjected to a leftward force. This can be seen by considering a transformation where the delta function potential barrier is quasi-statically pushed to the left: $\delta(q)\rightarrow \delta(q-q_0(t))$. Here, the displacement, $q_0(t)$, varies from 0 to $-\infty$ and satisfies the adiabaticity condition $|\dot{q}_0/q_0|\ll\omega$. The corresponding Hamiltonian is given by $\hat{H}_{\reflectbox{$\veryshortarrow$}}(t)=\frac{\hat{p}^2}{2m}+\frac{1}{2}m\omega^2 \hat{q}^2+\alpha\delta(\hat{q}-q_0(t))$ and the analysis of the problem now proceeds along the same lines as in Sec.~\ref{barrier}. (Notice that, if the quantum system had been projected onto the left side of the partition, we could have analogously defined a Hamiltonian $\hat{H}_{\veryshortarrow}$ describing the dynamics of a rightward quasi-static push of the barrier.) During the evolution, the eigenstates $|R_n\rangle$ contained in $\hat{\rho}_R$ remain instantaneous eigenstates of the Hamiltonian, with wave functions $\Psi_n^{q_0}$ satisfying the Schr\"odinger equation
\be
-\frac{\hbar^2}{2m}\frac{d^2}{dq^2}\Psi^{q_0}(q)+\frac{1}{2}m\omega^2q^2\Psi^{q_0}(q)+\alpha\delta(q-q_0)\Psi^{q_0}(q_0)=E^{q_0}\Psi^{q_0}(q)\,.
\label{instSchrodinger}
\ee
Recall that the strength of the potential barrier $\alpha$ is such that $\alpha\gg\sqrt{\hbar^3\omega/m}$. Therefore the wave functions $\Psi^{q_0}$ are only non-zero on the right side of the barrier and satisfy the boundary condition $\Psi^{q_0}(q_0)=0$.
The energy $E$ now depends on the position of the barrier $q_0$. Therefore, the boundary condition is seen to be an implicit equation relating $q_0$ and $E$ (see Eq.~\eqref{energydrift} in the appendix) and we plot its solutions in Fig.~\ref{barrierpush}.
\begin{figure}
      \includegraphics[width=0.4\textwidth]{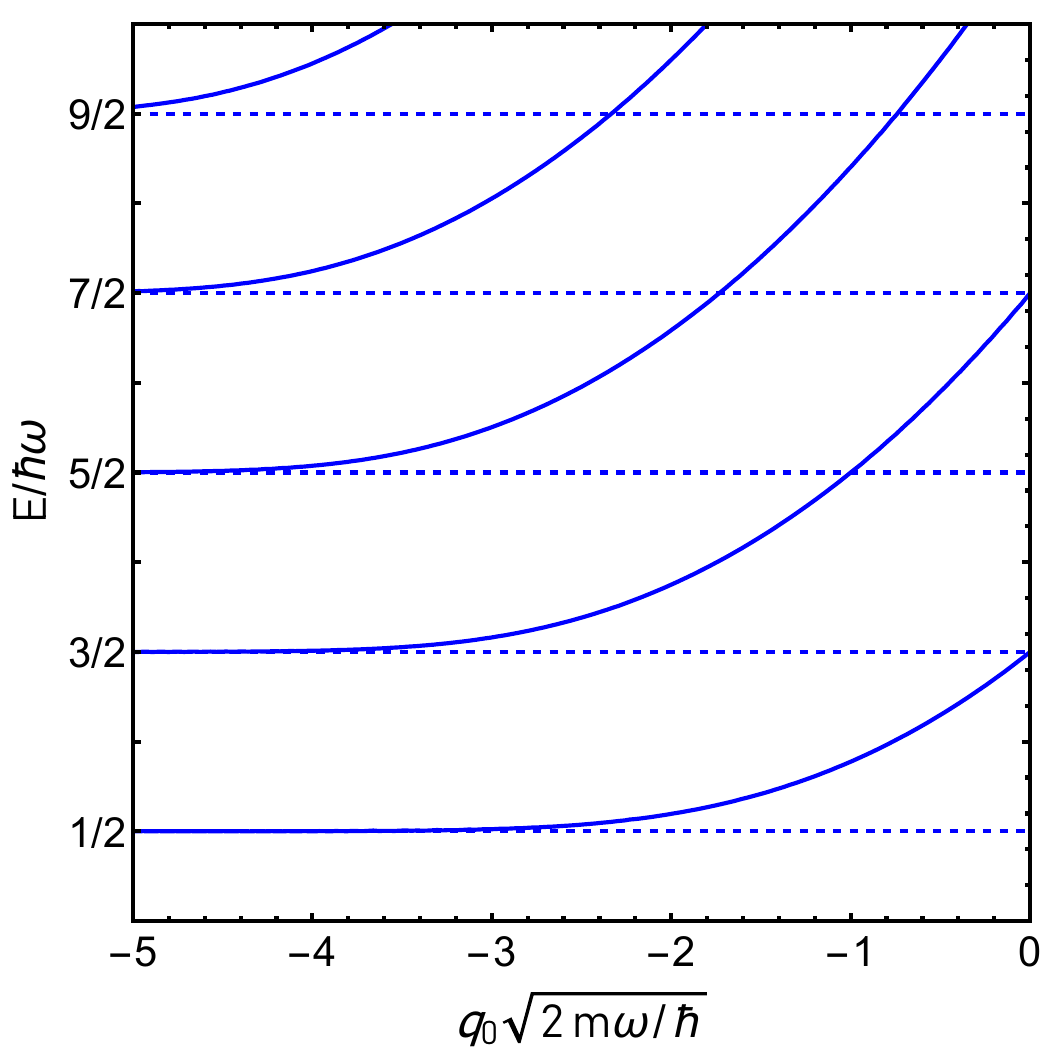}
 \caption{Graphical solution to the implicit equation relating $E$ and $q_0$, given by the Dirichlet boundary condition at the location of the barrier, $\Psi^{q_0}(q_0)=0$; the solid blue curve represents the energy levels of the right modes, $|R_n\rangle$, for a given value of $q_0$. The dashed blue lines show the energy levels of the harmonic oscillator without the barrier.}
\label{barrierpush}
\end{figure}
The force exerted on the barrier can now be computed semiclassically from the infinitesimal variation of the free energy of the system under an infinitesimal displacement of the barrier via
\be
F=-\frac{dA}{dq_0}\,.
\ee
(For an isothermal process, the first law of thermodynamics implies that the mechanical work done by the system is the opposite of the change in the free energy). This force is immediately seen to be negative, since for each energy eigenvalue branch in Fig.~\ref{barrierpush}, the energy decreases as the barrier moves to the left, 
and so does $A$. Indeed, $A=-\frac{1}{\beta}\ln Z=-\frac{1}{\beta}\ln \left(\sum_{n=0}^\infty e^{-\beta E_n}\right)$ is an increasing function of $q_0$. As expected, the localized particle exerts a pressure on the barrier which triggers the isothermal expansion of the right side of the well. In fact, the force at the beginning of the expansion can be calculated analytically (see Appendix \ref{appendixB}):
\be
F=-\sqrt{\frac{\hbar\omega^3m}{\pi}}\left(1-e^{-2\beta\hbar\omega}\right)^{-1/2}\,.
\label{initForce}
\ee
In principle, one could compute the force at every point of the expansion, but this would require a more extensive numerical treatment.

Moreover Fig.~\ref{barrierpush} shows that the states $|R_n\rangle$ evolve quasi-statically towards the eigenstates $|\psi_n\rangle$ of the simple harmonic oscillator as $q_0$ decreases from 0 to $-\infty$. The system returns to its initial state $\hat{\rho}_{\rm in}$ as it expands isothermally. The change in free energy during this process is
\be
A_{\rm in} - A_{R}= -\frac{1}{2}\hbar\omega-\frac{1}{\beta}\ln\left[\cosh\left(\frac{\beta\hbar\omega}{2}\right)\right]-\frac{1}{\beta}\ln 2\,.
\ee
This energy, supplied by the gas to the partition in the form of mechanical work, could be used by some agent, e.g. to lift an external load. The first two terms correspond to the work that the demon supplied to the system during the insertion of the barrier while the last term corresponds to extra energy that the demon is able to extract from the heat bath. This energetic net gain will accumulate with each subsequent cycle\cite{noPower} and seems to violate the second law of thermodynamics; in other words the quantum Szil\'ard engine appears to be a monothermal heat engine. We will see however that this is an artifact of the demon not being treated as a dynamical system in its own right.

\section{Saving the second law}
\label{secondlaw}

To see that the second law of thermodynamics is not violated during the quantum Szil\'ard engine cycle described above, we need to include the demon in the quantum analysis of the problem. Following Ref.~\onlinecite{2018PhR...755....1Z}, we model the demon by a two-state quantum pointer variable, i.e. a system whose states represent the possible positions of the display pointer of a measurement apparatus. The two states are denoted by $|D_L\rangle$ and $|D_R\rangle$, and correspond respectively to the demon measuring the particle to be on the left or right side of the harmonic well. We assume that the pointer variable is initially in the so-called neutral state $|D_0\rangle = \frac{1}{\sqrt{2}}\left(|D_L\rangle+|D_R\rangle\right)$ and therefore the state of the full 
$\{\text{particle}+\text{demon}\}$ system before barrier insertion is given by~\cite{endnote0bis}
\be
\hat{\rho}_{\rm in}\otimes |D_0\rangle\langle D_0|\,.
\ee

As the barrier is inserted quasi-statically, the full Hamiltonian is $\hat{H}_{\alpha}\otimes\hat{\mathds{1}}_D$ (where $\hat{\mathds{1}}_D$ is the identity operator on the pointer variable Hilbert space\cite{endnote1}) and at the end of barrier insertion the state of the system becomes
\be
\hat{\rho}_{\perp}\otimes |D_0\rangle\langle D_0|\,.
\ee

Up to this point, since the pointer and the quantum particle haven't interacted yet, there is no difference from the analysis in Sec.~\ref{cycle}. However, during the measurement phase, the pointer variable and the demon are coupled. The measurement interaction can be modeled via the following interaction Hamiltonian
\be
\hat{H}_{\rm int}=i\lambda \hat{\Pi}\otimes\left(|D_L\rangle\langle D_R|-|D_R\rangle\langle D_L|\right)\,,
\ee
which is turned on for a duration $\delta t=\frac{\pi\hbar}{4\lambda}$. Here $\lambda$ denotes the coupling strength between the pointer variable and the particle, and the observable $\hat{\Pi}$ is given by~\eqref{observable}. Since this Hamiltonian is time-independent, the evolution operator for the measurement process is simply given by
\be
\hat{U}_{\rm int}=e^{-\frac{i}{\hbar}\hat{H}_{\rm int}\delta t}=\frac{1}{\sqrt{2}}\left[\hat{\mathds{1}}_P\otimes \hat{\mathds{1}}_D +\hat{\Pi}\otimes \left(|D_L\rangle\langle D_R|-|D_R\rangle\langle D_L|\right)\right]\,,
\ee
where we have introduced the identity operator $\hat{\mathds{1}}_P$ on the particle Hilbert space and the last equality follows from expanding the exponential and using the properties $\hat{\Pi}^{2}=\hat{\mathds{1}}_P$ and $(|D_L\rangle\langle D_R|-~\!\!|D_R\rangle\langle D_L|)^2=-\hat{\mathds{1}}_D$. At the end of the measurement process, the state of the system is described by the density matrix
\be
\hat{U}_{\rm int}\, \left(\hat{\rho}_\perp\otimes |D_0\rangle\langle D_0|\right)\,\hat{U}^\dag_{\rm int} = \frac{1}{2}\left(\hat{\rho}_L\otimes |D_L\rangle\langle D_L|+\hat{\rho}_R\otimes |D_R\rangle\langle D_R|\right)\,.
\ee
We see that, as a result of the interaction between the particle and the pointer variable, the two systems become correlated. Indeed, if the particle is on the left side of the well then the state of the  pointer variable is necessarily $|D_L\rangle$, and similarly if the particle is on the right, the state is $|D_R\rangle$. This corresponds exactly to what we would expect from a strong, or projective, quantum measurement. This can be made more precise by considering the reduced entropy of the pointer variable, i.e. the entropy computed from the reduced density matrix, obtained by performing a trace of the full density matrix over the possible states of the particle. Before the measurement, the reduced density matrix for the pointer variable is simply the pure state $|D_0\rangle\langle D_0|$ and therefore its reduced entropy is 0. After the measurement, however, the reduced density matrix becomes $\frac{1}{2}\left(|D_L\rangle\langle D_L|+|D_R\rangle\langle D_R|\right)$ and the corresponding reduced entropy grows by $k_B\ln 2$. (Notice however that the reduced entropy of the quantum particle does not change during the measurement process, which is a sign that the {\it mutual information} is non-zero.\cite{2018PhR...755....1Z})

During the final step of the cycle, {\it i.e.} the quasi-static isothermal expansion of the one-particle quantum gas, the full Hamiltonian is $\hat{H}_{\veryshortarrow}\otimes |D_L\rangle\langle D_L|+\hat{H}_{\reflectbox{$\veryshortarrow$}}\otimes |D_R\rangle\langle D_R|$. Here we have used the quasi-statically-varying Hamiltonians $\hat{H}_{\veryshortarrow}$ and $\hat{H}_{\reflectbox{$\veryshortarrow$}}$ that describe the dynamics of the barrier being slowly pushed to the right or to the left, which were introduced in Sec.~\ref{joule}. Denoting by $\hat{U}_{\veryshortarrow}$ and $\hat{U}_{\reflectbox{$\veryshortarrow$}}$ the associated unitary time evolution operators, the state of the system during expansion is
\be
\frac{1}{2}\left(\hat{U}_{\veryshortarrow}\hat{\rho}_L\hat{U}^\dag_{\veryshortarrow}\otimes |D_L\rangle\langle D_L|+\hat{U}_{\reflectbox{$\veryshortarrow$}}\hat{\rho}_R\hat{U}^\dag_{\reflectbox{$\veryshortarrow$}}\otimes |D_R\rangle\langle D_R|\right)\,.
\ee
As the barrier is pushed all the way to infinity, $\hat{U}_{\veryshortarrow}\hat{\rho}_L\hat{U}^\dag_{\veryshortarrow}=\hat{U}_{\reflectbox{$\veryshortarrow$}}\hat{\rho}_R\hat{U}^\dag_{\reflectbox{$\veryshortarrow$}}=\hat{\rho}_{\rm in}$ and the final state of the system is
\be
\frac{1}{2}\hat{\rho}_{\rm in}\otimes\left(|D_L\rangle\langle D_L|+|D_R\rangle\langle D_R|\right)\,.
\ee
It is clear from this expression that while the quantum particle is returned to its initial state at the end of the cycle, the pointer variable remains in a mixed state. To be able to restart the cycle, one would need to “purify" the state of the pointer variable and reset it to its neutral state $|D_0\rangle$. For example, an external observer could measure the state of the demon, thereby projecting the demon into one of the eigenstates $|D_L\rangle$ or $|D_R\rangle$, following which a unitary transformation could be used to ‘rotate’ the state to $|D_0\rangle$. By the first law of thermodynamics, whatever strategy is used would unavoidably incur a cost of at least $k_B T \ln 2$ of free energy, and a corresponding entropy increase in the wider environment.\cite{endnote2} The resetting operation thus negates the apparent benefits of the Szilard engine cycle and saves the second law of thermodynamics.

\section{Discussion}
\label{discussion}

We have described in detail how a Szil\'ard engine would work for a quantum particle confined to a harmonic well. Unlike the more familiar case of the infinite square well, all the thermodynamic functions could be solved for analytically at the end of each stage of the cycle, allowing us to see the behavior of the engine at both high and low temperatures. We then showed how the second law of thermodyamics is apparently violated when the Maxwell demon operating the engine is not treated dynamically. Seemingly, the first engine cycle extracts an amount of useful work equal to $k_BT\ln 2$ from the thermal bath. However, a full quantum analysis showed explicitly that a dynamical quantum demon is unable to operate the engine for more than one cycle, and that its entropy increases by $k_B\ln 2$ as it recovers information about the position of the particle. To continue the cycle, the demon would need to be reset (thus lowering its entropy), which requires at least as much free energy as is available from the engine cycle.

\begin{figure}[t]
\includegraphics[width=0.6\linewidth]{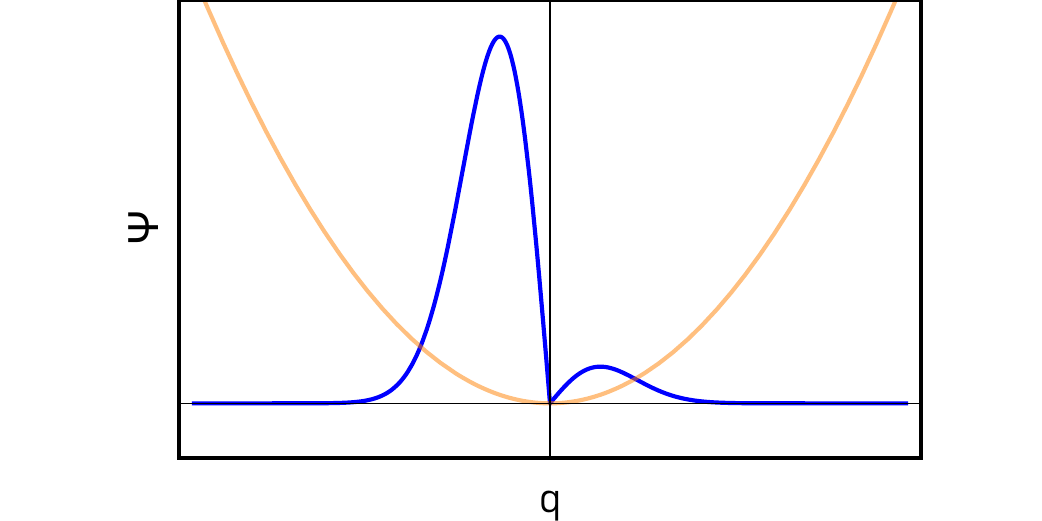}
\caption{After a weak measurement, the wave function (blue) is partially projected into the left state. A piston is then inserted and there is a pressure that pushes the piston rightward. This pressure is counteracted by a pressure exerted on the right side of the piston by the ``leftover" wave function getting pushed up the potential (pictured in orange).}
\label{WMP}
\end{figure}

An interesting extension of this work would be to study how the quantum Szil\'ard engine operates when the measurement process is weak i.e. non-projective.\cite{WeakMeasurement} In the crude measurement model described in Sec.~\ref{secondlaw}, this occurs presumably when $\frac{\delta t \lambda}{\hbar}\neq \frac{\pi}{4}$ and the interaction with the pointer variable doesn't completely project the quantum state to one side of the barrier (see Fig.~\ref{WMP}).  
It is then unclear to what extent the results of the present paper carry over. An ensemble of quantum Szil\'ard engines and demons would be required to reliably extract any useful work, as there would be a non-zero probability for any single demon to inaccurately determine the location of the particle and therefore to install the piston in an orientation that cannot extract energy. Also, without fully projecting the particle onto one side of the well or the other, the isothermal expansion during the last stage of the thermodynamic cycle would be incomplete, and the amount of useful work extracted during the first cycle would be less than in the case of a strong measurement. These generalizations suggest that one should define `weak information' and average over the ensemble to investigate the trade-off between average work performed and weak information garnered and erased. We leave the extension of the thermodynamic analysis to this weak domain to future investigation.

\section*{Acknowledgements}

We should like to thank S.~M. Carroll and M.~E. Tegmark for helpful comments. During this work LT and GZ were supported by the Foundational Questions Institute (FQXi). LT is also supported by the U.S. Department of Energy, under grant number DE-SC0019470. GZ also acknowledges financial support from {\it Moogsoft} and the Archelon Time Science Fellowship.

\appendix
\section{Energy levels of the even modes}
\label{appendixA}
We study the solutions $\Psi^\alpha_n$ and $E^\alpha_n$ of Eq.~\eqref{instchrodinger} for even $n$. First note that, because of the symmetry, we need to solve Eq.~\eqref{instchrodinger} only for $q>0$. In this domain, the delta function does not contribute and the equation reduces to the Weber differential equation.\cite{G&R,mathSite} There are, in general, two linearly-independent solutions to this equation, but only one is bounded as $q\rightarrow\infty$,
\be
\Psi^\alpha(q) = ND_a\left(q\sqrt{\frac{2m\omega}{\hbar}}\right)\,,
\label{parabolic}
\ee
where $a = \frac{2E-\hbar\omega}{2\hbar\omega}$, and $D_a$ are the so-called parabolic cylinder functions.\cite{G&R,mathSite} The normalization of the wave function is
\be
N = \left(\sqrt{\frac{\pi\hbar}{4m\omega}}\,\frac{\psi\left(\frac{1-a}{2}\right)-\psi\left(-\frac{a}{2}\right)}{\Gamma(-a)}\right)^{-1/2}\,,
\ee
where $\Gamma(x)$ is the gamma function, and $\psi(x)$ is the digamma function. (For the usual simple harmonic oscillator energies, \eqref{parabolic} recovers the familiar Hermite polynomials.\cite{mathSite}) To extract the even state energy levels, we need to use the fact that the delta function piece of the potential sets a condition on the behavior of the wave function at $q=0$
, which manifests itself as a discontinuity of $\frac{d\Psi^\alpha}{dq}$. Integrating Eq.~\eqref{instchrodinger} from $q=-\epsilon$ to $q=+\epsilon$ and taking the limit $\epsilon\rightarrow 0$ yields
\be
\frac{d}{dq}\Psi^\alpha(0_+) = \frac{m\alpha\Psi^\alpha(0)}{\hbar^2}\,.
\ee
This gives the quantization condition
\be
D'_a(0)\sqrt{\frac{2\hbar^3\omega}{m}}=\alpha D_a(0)\,,
\label{quantCondition}
\ee
where\cite{ParCylD}
\ba
D_a(0) &=& \frac{2^{\frac{2E-\hbar\omega}{4\hbar\omega}}\sqrt{\pi}}{\Gamma\left[\frac{1}{2}-\frac{2E-\hbar\omega}{4\hbar\omega}\right]}\,,\\
D'_a(0) &=& -\frac{2^{\left(\frac{2E+\hbar\omega}{4\hbar\omega}\right)}\sqrt{\pi}}{\Gamma\left[-\frac{2E-\hbar\omega}{4\hbar\omega}\right]}\,.
\ea
Eq.~\eqref{quantCondition} then yields Eq.~\eqref{quantization} and may be solved numerically for the energy levels of the even states, for a given value of $\alpha$.

\section{Initial force on the barrier during isothermal expansion}
\label{appendixB}
We consider the change in the states $|R_n\rangle$ after a small displacement of the barrier to the left. Their wave functions $\Psi^{q_0}_n$ and corresponding energy eigenvalues $E^{q_0}_n$ satisfy Eq.~\eqref{instSchrodinger}, subject to a Dirichlet boundary condition at $q=q_0$. More precisely, 
\be
\Psi^{q_0}(q)\propto D_a\left(q\sqrt{\frac{2m\omega}{\hbar}}\right)
\ee 
where $a = \frac{2E-\hbar\omega}{2\hbar\omega}$ and the boundary condition reads
\be
D_a\left(q_0\sqrt{\frac{2m\omega}{\hbar}}\right)=0\,.
\label{energydrift}
\ee

We now derive the initial force exerted by the particle on the barrier during the isothermal expansion, given in the main text in Eq.~\eqref{initForce}, assuming that the barrier is sufficiently massive that its own quantum effects may be neglected. Recall that at constant temperature, the force can be obtained from the free energy, $A$, by
\begin{equation}
F=-\frac{dA}{dq_0},
\label{appForce}
\end{equation}
where $q_0$ is the position of the barrier. For concreteness, we will assume that the demon measures the particle to be on the right side of the barrier. The free energy at the beginning of the isothermal expansion is given by
\begin{equation}
A_R = -\frac{1}{\beta}\ln Z_R = -\frac{1}{\beta}\ln\left(\sum_{n=0}^{\infty}e^{-\beta E_n}\right),
\end{equation}
where $E_n=\hbar\omega(2n+3/2)$ are the energy levels of the energy eigenstates $|R_n\rangle$. Taking the derivative as in Eq.~\eqref{appForce},
\begin{equation}
-\frac{dA_R}{dq_0}=\frac{1}{\beta Z_R}\frac{dZ_R}{dq_0}=-\frac{1}{Z_R}\sum_{n=0}^{\infty}e^{-\beta E_n}\frac{dE_n}{dq_0}.
\label{appdAdq}
\end{equation}

In order to compute the above sum, we consider a small displacement $\delta q_0$ of the barrier to the left. Expanding the boundary condition \eqref{energydrift} to linear order in $\delta q_0$, we find
\begin{equation}
0 = D_a(0) + \left(\sqrt{\frac{2m\omega}{\hbar}}D'_a(0)+\frac{dE_n}{dq_0}\frac{1}{\hbar\omega}\frac{d}{da}D_a(0)\right)\delta q_0 + \mathcal{O}(\delta q_0{}^2)
\end{equation}
which implies
\begin{equation}
\frac{dE_n}{dq_0} = -\sqrt{2m\hbar\omega^3}\frac{D'_a(0)}{\frac{d}{da}D_a(0)}\,.
\label{appdEdq}
\end{equation}
The two derivatives on the right-hand side of Eq.~\eqref{appdEdq} are given by~\cite{ParCylD}
\begin{align}
D'_a(0) &= -\frac{2^{\frac{1+a}{2}}\sqrt{\pi}}{\Gamma\left(-\frac{a}{2}\right)}\,,\\
\frac{d}{da}D_a(0) &= \frac{2^{-1+\frac{a}{2}}\sqrt{\pi}}{\Gamma\left(\frac{1-a}{2}\right)}\left(\ln2+\psi\left(\frac{1-a}{2}\right)\right)\,,
\end{align}
where $\psi(x)$ is the digamma function. We can now write Eq.~\eqref{appdEdq} as
\begin{equation}
\frac{dE_n}{dq_0} = \sqrt{m\hbar\omega^3}\frac{4\Gamma\left(\frac{3\hbar\omega-2E_n}{4\hbar\omega}\right)}{\Gamma\left(\frac{\hbar\omega-2E_n}{4\hbar\omega}\right)\left[\ln2+\psi\left(\frac{3\hbar\omega-2E_n}{4\hbar\omega}\right)\right]}.
\end{equation}
Note that the numerator and denominator in the above equation diverge as $E_n\rightarrow\hbar\omega(2n+3/2)$, so we must consider the limiting value. In this limit, the expression simplifies considerably:
\begin{equation}
\frac{dE_n}{dq_0} = \sqrt{\frac{m\hbar\omega^3}{\pi}}\frac{(2n+2)!}{4^nn!(n+1)!}\,.
\end{equation}
It can be shown, by the Taylor expansion of the binomial series, that
\begin{equation}
\frac{2}{(1-x)^{3/2}} = \sum_{n=0}^{\infty}\frac{(2n+2)!}{4^nn!(n+1)!}x^n\,.
\end{equation}
Therefore, the sum in Eq.~\eqref{appdAdq} may be evaluated, using also the expression for $Z_R$ in Eq.~\eqref{ZLR},
\begin{equation}
F=-\frac{dA}{dq_0}= -2\sqrt{\frac{m\hbar\omega^3}{\pi}}\left(1-e^{-2\beta\hbar\omega}\right)^{-1/2}.
\end{equation}


\begin{thebibliography}{99}
\bibitem{ShannonEntropy} The Shannon entropy is defined without the factor of Boltzmann's constant, but here it is required to obtain the correct units.
\bibitem{noPower} Although there is energy output from the engine, the quasi-static assumption implies zero net power. This is a familiar idealization in thermodynamics, as the Carnot engine also evolves quasi-statically during its cycle. Relaxing this quasi-static condition reduces the efficiency of the engine.
\bibitem{endnote1} Technically $\hat{\mathds{1}}_D$ should be replaced by the quantum Hamiltonian of the pointer variable. However we neglect the quantum dynamics of the pointer variable in this work. This can safely be done as long as the two energy eigenvalues corresponding to $|D_L\rangle$ and $|D_R\rangle$ are separated by less than $\hbar/\tau$ (where $\tau$ is the characteristic time scale of the problem).
\bibitem{endnote2} Of course, one could consider a quantum demon at a temperature $T_D<T$, in which case the energy required to ``reset" the demon is less than the amount extracted from one engine cycle. However in this case the engine is no longer monothermal, so there is no violation of the second law of thermodynamics.
\end{thebibliography}
\end{document}